\documentclass{article}
\usepackage[utf8]{inputenc}

\usepackage{enumerate}
\usepackage{hyperref}
\usepackage{graphicx}
\usepackage{amsmath}
\usepackage{amsfonts}
\usepackage{amssymb}
\usepackage[none]{hyphenat}
\usepackage{float}
\usepackage{authblk}
\usepackage{titlesec}
\usepackage[top=2.5cm, bottom=2.5cm, left=3cm, right=3cm]{geometry}

\begin{document}
\title{\textbf{Bayesian Strategies for Likelihood Ratio Computation in Forensic Voice Comparison with Automatic Systems}}
%\author[XX]{XX}
\author[]{Daniel Ramos}
\author[]{Juan Maroñas}
\author[]{Alicia Lozano-Diez}
\affil[]{AUDIAS Lab (Audio, Data Intelligence and Speech) \\ http://audias.ii.uam.es \\ Escuela Politecnica Superior, Universidad Autonoma de Madrid}
\affil[]{\texttt{daniel.ramos@uam.es,jmaronasm@gmail.com,alicia.lozano@uam.es}}
\date{September, 2017\footnote{This work has been published in Subsidia: Tools and Resources for Speech Sciences in June 2017}}

%  {[}LOGOTIPO{]} \hfill {[}{]}
  
%  \hfill {[}{]}
  
%  \hfill {[}{]}
  
    \maketitle
\fontsize{8.5pt}{10pt}

\noindent
% \textbf{Citation:} {[}Autor/autores{]}. {[}(Año){]}. {[}Título{]}. {[}Publicación{]}. {[}On line{]}.

\bigskip

\fontsize{9pt}{10.5pt}

\textsc{\textbf{Abstract:}}

\noindent This paper explores several strategies for Forensic Voice Comparison (FVC), aimed at improving the performance of the LRs when using generative Gaussian score-to-LR models. First, different anchoring strategies are proposed, with the objective of adapting the LR computation process to the case at hand, always respecting the propositions defined for the particular case. Second, a fully-Bayesian Gaussian model is used to tackle the sparsity in the training scores that is often present when the proposed anchoring strategies are used. Experiments are performed using the 2014 i-Vector challenge set-up, which presents high variability in a telephone speech context. The results show that the proposed fully-Bayesian model clearly outperforms a more common Maximum-Likelihood approach, leading to high robustness when the scores to train the model become sparse.

\bigskip

\bigskip

\fontsize{10pt}{12pt}

\section{\textsc{Introduction}}
In forensic voice comparison (FVC) using automatic systems, a score is typically transformed into a likelihood ratio (LR) by using some probabilistic model \cite{gonzalez07IEEETrans,drygajlo16forensicSpeakerRecognition}. Recently, this methodology has been proposed as the recommended way of reporting in court using automatic speaker recognition systems, in the context of the speech and audio laboratories belonging to the European Network of Forensic Science Institutes (ENFSI) \cite{drygajlo16methodologicalGuidelinesENFSI}. Also, this is the typical strategy for LR computation in other biometric systems, where a score is the usual output when comparing two biometric specimens \cite{ramos16forensicBiometrics}. This approach can be also used in general when a model or method previously computes a score among two evidential materials, such as two handwritten documents or two chemical profiles \cite{hepler11handwritingAnchoring,bolck15scoreFeatureLRMDMA}.

Many techniques have been proposed in the past to do this score-to-LR transformation. Perhaps the most straightforward way is to assign a probability distribution \footnote{In a domain of continuous scores, it will be a probability density function; and a probability mass function if the scores are discrete} to the score from the automatic system, given each one of the propositions in the forensic case. As score normalization techniques with Gaussianization properties are frequently an intrinsic stage of automatic speaker recognition systems\cite{navratil03awetnorm}, the use of Gaussian distributions appears as a sensible choice\cite{brummer14BayesianForensicReporting}. Anyway, probabilistic distributions are typically assigned to scores using training strategies like Maximum Likelihood (ML) or Maximum a Posteriori (MAP) \cite{alberink14fingermarkConditioning,ramos06oddyseyMAP}. Another popular approach is logistic regression, also trained with ML or MAP \cite{brummer07fusion}, because of its good robustness properties over the whole score range, and the possibility of regularization, the latter affecting the calibration of the final LRs negatively. Recently, fully-Bayesian strategies have been proposed, in order to cope with the sparsity in the amount of scores available for training, while showing good calibration performance \cite{brummer14BayesianForensicReporting}.

Another main problem in FVC is the use of the specimens and databases to compute a better LR in a given case. It is well known that speech is extremely variable according to many conditions (such as phonetic content, acoustic environment, emotional state, transmission channel and so on), and it has been also observed that this variability affects the score range quite seriously \cite{perez10Interspeech,mandasari13qualityBasedCalibration}. Therefore, in order to compute a LR that presents good performance for a given case, two actions may be considered. First, the automatic speaker recognition system must implement powerful session variability compensation techniques in order to reduce the variability of the scores to be transformed into LRs. Second, the LR model itself must take into account the conditions of the speech in each case, while respecting the propositions and conditioning information of the case itself, because the distribution of the training scores have to fit the distribution of the score in the case. The LR model can address the latter by the selection of the data to compute the scores used to train the LR model. This data selection process and the subsequent strategy to compute the scores has been dubbed \emph{anchoring} \cite{hepler11handwritingAnchoring,alberink14fingermarkConditioning}. In this paper, we will propose some anchoring strategies, showing experiments that support their use in challenging FVC scenarios.

An additional problem with some \emph{anchoring} strategies is that they typically results in a very small amount of scores to train the LR model. Unfortunately, these schemes are the most typically proposed ones in FVC \cite{drygajlo16forensicSpeakerRecognition}. The sparsity in the amount of scores typically affects the likelihood of the prosecution proposition, in the numerator of the LR, because the available data from a given suspect is often very limited in a case. In order to overcome this problem, we will test the use of a Gaussian fully-Bayesian model to cope with the uncertainty due to data sparsity.

Thus, the contribution of this work is two-fold: first, different anchoring schemes are tested, to show the adequacy of taking into account the conditions of the case in the anchoring process. Second, fully-Bayesian Gaussian models are proposed as an efficient way of reducing the problem of data sparsity in the training scores. Both experimental contributions are set-up in a highly challenging scenario of speech variability under telephone conditions: the NIST 2014 i-Vector Challenge.

The paper is organized as follows: Section \ref{sec__SpeechHandling} describes the anchoring schemes used in this paper, and their motivation. Section \ref{sec__models} will introduce the models to be compared in this article, namely Gaussian ML and Gaussian Fully-Bayesian. Finally, Section \ref{sec__experiments} describes the database and experimental protocol in the highly variable context of the 2014 NIST i-Vector Challenge competition; and shows the results supporting the research hypothesis. Conclusions are finally drawn in Section \ref{sec__conclusions}.

\section{\textsc{Handling Speech Data in FVC using Automatic Systems}}
\label{sec__SpeechHandling}

\subsection{Propositions in FVC}
A FVC case has the following typical elements. On the one hand, there are some questioned speech materials (namely, one or more questioned recordings) of disputed origin, also known as \emph{trace}, and in general of an incriminatory nature. On the other hand, a suspect identified on the basis of other information, and from whom some control materials are recorded, namely one or more \emph{reference} or \emph{control} recordings. The problem consists in expressing the value of such evidence with respect to two propositions defined in the case. In this context, the identity of the suspect is typically known, as well as some other features that can be extracted from contextual information in the case, or from the speech materials themselves. This motivates the following typical definition of the propositions in FVC:

\begin{itemize}
    \item $H_p$: the questioned materials come from the suspect the suspect.
    \item $H_d$: the questioned materials do not come from the suspect, but another individual from a given population of potential sources.
\end{itemize}

Note that $H_p$ and $H_d$ differ in the assumption of the origin of the trace $q$, which is unknown in the case. Therefore, conditioning in the propositions will imply a change of the source of the trace in the generation of training scores. Also, it is assumed that the suspect, and no other possible individual, has originated the reference materials in both cases, because the suspect is referred to in both propositions themselves. Such propositions have been dubbed \emph{source-specific} or, if the potential source is a person, respectively \emph{person-specific} \cite{ramos16forensicBiometrics}. They are the most common in FVC, where a suspect of known identity has been usually accused by a court of law.

We define $H$ as the random variable representing the proposition, with alphabet $\left\{ H_p, H_d \right\}$. Since the score generated by the comparison between the questioned and control materials must be evaluated in the context of those competing propositions, the likelihood ratio (LR) formula naturally arises \cite{gonzalez07IEEETrans,drygajlo16methodologicalGuidelinesENFSI,drygajlo16forensicSpeakerRecognition}:

\begin{equation}
    LR = \frac{p\left( \left. s_c \right| H_p\right)} {p\left( \left. s_c \right| H_d\right)}
\label{eq__LR}
\end{equation}

\noindent where $s_c$ is the score, \emph{observed} from the comparison of the questioned and control materials, as given by the automatic system; and $p\left( \left. . \right| . \right)$ denotes a conditional probability density. In this way, the value of the evidence is given by the LR, and can be reported according to common procedures \cite{drygajlo16methodologicalGuidelinesENFSI}.

\subsection{Generating scores: \emph{anchoring}}

In order to assign the densities in \ref{eq__LR}, two sets of training scores must be generated. Each training score will be accompanied by a \emph{class label}, where classes are the $H_p$ and $H_d$ values from variable $H$ in this forensic scenario. A training score $s_i$ and its corresponding class label $H_i$ are represented as $(s_i,H_i)$. The total set of training scores will be $S=\left\{ S_p,S_d\right\}$, where $S_p = \left\{ (s_p^{(1)},H_p)),...,(s_p^{(N_p)},H_p)\right\}$ for the density in the numerator (conditioned to $H_p$) and $S_d = \left\{ (s_d^{(1)},H_d),...,(s_d^{(N_d)},H_d)\right\}$ 
for the density in the denominator (conditioned to $H_d$). The process for speech data selection and score generation of training scores is known as \emph{anchoring} \cite{hepler11handwritingAnchoring,alberink14fingermarkConditioning}, and must consider the following facts:

\begin{enumerate}
\item The definition of the propositions will define how the scores $S_p$ and $S_d$ must be generated.

\item The high variability of the speech signal previously mentioned seriously compromise the distribution of the scores used to train LR models \cite{perez10Interspeech,mandasari13qualityBasedCalibration}. Therefore, the generation of the scores must be done in the most similar speech conditions as those in the case, otherwise the densities will not represent the observed score $s_c$ in the case at hand \cite{drygajlo16forensicSpeakerRecognition}.
\end{enumerate}

Let $s_c$ be generated from the comparison of the questioned speech $q$ and the reference speech $r$. Thus, $s_c = \Delta\left(q,r\right)$, where $\Delta\left(.,.\right)$ represents the score computation algorithm of the automatic speaker recognition system\footnote{Although the $\Delta$ notation might suggest similarity, scores in speaker recognition systems most often also include population models and typicality. In fact, i-Vector PLDA systems output log-likelihood-ratios, but presenting poor calibration \cite{matejka11PLDA}}. Also, it is assumed that $\Delta\left(q,r\right) = \Delta\left(r,q\right)$, as it happens with many speaker recognition systems\cite{matejka11PLDA} based on i-Vectors. Then, from both facts described above, some conclusions can be extracted for a FVC case using automatic systems:

\begin{itemize}
    \item In order to satisfy $H_p$, for the density in the numerator of the LR, the training scores must be generated using traces from the suspect, known as \emph{speech controls} of \emph{pseudo-traces}; to be compared with reference speech also from the suspect. Moreover, in FVC a \emph{person-specific} proposition implies that using same-person scores coming from other individuals different from the suspect will not be adequate, since the distribution of scores of different individuals is known to be highly variable \cite{doddington98zoo}. Therefore, comparing speech controls $\left\{q_p^{(i)}\right\}$ to other reference speech materials from the suspect, namely $\left\{r_p^{(j)}\right\}$, $S_p = \left\{ \Delta \left( q_p^{(i)},r_p^{(j)}\right) \right\}$ is obtained, with $\left|S_p\right| = N_p$.
    \item In order to fit the conditions of the case, both the trace and the reference speech used to generate training scores should be selected according to the following criteria:
    \begin{itemize}
        \item Reference speech: As the identity of the suspect is known to be the one in $r$, $\left\{r_p^{(j)}\right\}$ should consist of speech data from the suspect in conditions as close as possible to the conditions of $r$. In the limit, the best possible fitting is $r_p^{(j)}=r$, and to use a number $N_p$ of speech controls $q_p^{(i)}$.
        \item Trace: The speech controls $q_p^{(i)}$ must have conditions as close as possible to $q$. Otherwise, the model obtained using $S_p$ will not represent the model that could generate $s_c=\Delta(q,r)$ if $r$ and $q$ come from the same source, because a change in the conditions of the trace $q$ will most probably affect the score distribution.
    \end{itemize}
    \item In order to satisfy $H_d$, for the density in the denominator of the LR, the training scores must be generated using traces coming from individuals being potential origins of the trace. In forensic practice, these potential origins are assumed to be drawn from a so-called \emph{population} of individuals. Also, since the definition of the proposition is \emph{person-specific}, it is assumed that the suspect, and no other individual, has generated the reference speech recordings in the case. Therefore, the scores $S_d$ must be generated by traces $q_d^{(i)}$ generated by individuals from the population of potential origins, compared to reference speech segments $r_d^{(j)}$ from the suspect. Thus, $S_d = \left\{ \Delta \left( q_d^{(i)},r_d^{(j)}\right) \right\}$, with $\left|S_d\right| = N_d$.
    \item In order to fit the conditions of the case, the same rationale as for $S_p$ applies. Thus:
    \begin{itemize}
        \item Reference speech: As the identity of the suspect is known to be the one in $r$, the best fitting in the conditions of $\left\{r_d^{(j)}\right\}$ is to use speech data from the suspect in conditions as close as possible to the conditions of $r$. In the limit, the best possible fitting is to always use $r_d^{(j)}=r$, and a number $N_d$ of traces $q_d^{(i)}$.
        \item Trace: The $q_d^{(i)}$ traces must have conditions as close as possible as in $q$. Otherwise, the model obtained using $S_d$ will not represent the model that could generate $s_c=\Delta(q,r)$ if $q$ comes from other individual than the suspect.
    \end{itemize}
\end{itemize}

According to the aforementioned conclusions, in this work we propose two anchoring schemes:

\begin{enumerate}
\item \emph{Suspect-anchored} (SA): 
\begin{itemize}
    \item Scores in $S_p = \left\{ \Delta \left( q_p^{(i)},r_p^{(j)}\right) \right\}$ are anchored to the speaker, and therefore $\left\{q_p^{(i)}\right\}$ are speech controls from the suspect in conditions as close as possible to the ones in $q$, and $\left\{r_p^{(j)}\right\}$ are speech segments from the suspect in conditions as close as possible to the ones in $r$. 
    \item Scores in $S_d = \left\{\Delta\left( q_d^{(i)},r_d^{(j)}\right) \right\}$ are also anchored to the speaker, and therefore $\left\{q_d^{(j)}\right\}$ will be speech segments from other speakers from the population of potential origins, in conditions as close as possible to the ones in $q$; and $\left\{r_d^{(j)}\right\}$ will be speech segments from the suspect in conditions as close as possible to the ones in $r$. In fact, $\left\{r_p^{(j)}\right\}$ = $\left\{r_d^{(j)}\right\}$.
\end{itemize}
\item \emph{Reference-anchored} (RA):
\begin{itemize}
    \item Scores in $S_p = \left\{ \Delta \left( q_p^{(i)},r\right) \right\}$ are anchored to the reference speech $r$, and therefore $\left\{q_p^{(i)}\right\}$ are speech controls from the suspect in conditions as close as possible to the ones in $q$, and $r_p^{(j)}=r$. 
    \item Scores in $S_d = \left\{\Delta\left( q_d^{(i)},r\right) \right\}$ are also anchored to the reference speech $r$, and therefore $\left\{q_d^{(j)}\right\}$ will be speech segments from other speakers from the population of potential origins, in conditions as close as possible to the ones in $q$; and $r_d^{(j)}=r$\footnote{As described, the following proposed Suspect-Anchoring and Reference-Anchoring schemes make a distinction about how the scores are generated, not a distinction in the kind of statistical model that it used}.
\end{itemize}

\end{enumerate}

One of our research hypotheses is that RA will offer better performance than SA, since the fitting to the conditions in $r$ is the most perfect one for RA. Also, as it can be seen, the use of these anchoring schemes may lead to a substantially high number of scores $N_d$ for assigning the denominator of the LR, because $\left\{q_d^{(i)}\right\}$ can be found to be large. However, $N_p$ can be very low, because in FVC the amount of data from the suspect is usually very limited. Therefore, data sparsity must be addressed mainly for $S_p$.

\section{\textsc{Generative Models for LR computation from scores}}
\label{sec__models}
Computing the LR can be done following discriminative and generative approaches. An example of discriminative approach is logistic regression \cite{brummer07fusion,morrison13TutorialLogisticRegression}, widely used in automatic speaker recognition. However, generative approaches have been recently proposed as a robust alternative to other methods \cite{brummer14BayesianForensicReporting}. Reasons are given below.

% Juan, he quitado lo que sigue porque en el mundo forense crea una gigantesca confusión. Todo el mundo usa logistic regression, pero cuando he intentado contarles que se calcula el posterior y extraemos el prior, se vuelven locos. Mejor no mencionarlo, y que se miren las referencias.

%where we train the model for computing $P(c|s)$ where $c$ represents our target (a class of interst, such as $H_p$ or $H_d$ in our case) and $s$ represent the score. From this posterior probability and following the Bayes rule we can extract the $\log{LR}$ as a function of the score.\\

In this work we follow two generative approaches, where the $\log{LR}$ is computed from the ratio between the likelihood distributions, following equation \ref{eq__LR}.

%\begin{equation}
%    \log{LR}=\log{\frac{p(s|c_1)}{p(s|c_2)}}
%    \label{equ:1}
%\end{equation}

These likelihood distributions are computed from the joint probability density $p(s,H)$ or more exactly from the model representing the joint probability density $\widehat{p}(s,H|\theta)$ where for convenience we will replace $\widehat{p}$ by $p$ and $\theta$ represents the parameters of the model as a vector. In this work we implement these models using Maximum Likelihood (ML) and Bayesian Inference (BI).

\subsection{Maximum Likelihood}

Let $S=\{(s^{(i)},H_i)\}^N_{i=1}$ be a dataset drawn i.i.d. from the model distribution, that is, each sample is independent from the samples in the same class and from the samples in the other class (does not provide information about the other class). Thus, the likelihood function is defined as:\\
\begin{equation}
    p(S|\theta)=\prod^N_{i=1} p(s_i|H_i,\theta)\cdot p(H_i)
\end{equation}
    
One can show that under this condition we can optimize each term of the likelihood function: $p(x|H_p)$,$p(x|H_d)$,$p(H_p)$ and $p(H_d)$; independently. For an example see \cite{Bishop:2006:PRM:1162264} section 4.2.2. For the task we address we are only interested in estimating the parameters of the likelihood function, because the prior probabilities are not the responsibility of the forensic evaluation process, and we can express our model thus:

\begin{equation}
    p(S=\{s^{(1)},s^{(2)},..s^{(N)}\}|\theta)=\prod^N_{i=1} p(s_i|\theta)
    \label{equ:fact}
\end{equation}

Setting the derivative to zero and finding the maximum value we end up with a choice of the parameters. For a univariate Gaussian model, $\theta=(\mu_j,\sigma_j^2)$ with $j \in \{p,d\}$, as:

\begin{align}
    \mu_j &= \frac{1}{N_j}\underset{i \in N_j}{\sum}s_j^{(i)}\\
    \sigma_j^2 &= \frac{1}{N_j-1}\underset{i \in N_j}{\sum}(s_j^{(i)}-\mu_j)^2
\end{align}

\subsection{Bayesian Parameter Inference}

The approach of Bayesian inference starts by assuming we want to assign the predictive distribution $p(s|S)$, that is a distribution over the random variable $s$ whose form depends directly on the data set and not on a set of parameters. The key idea is marginalization over all the possible parameters of the underlying model. Hence:

\begin{equation}
    p(s|S)=\int_{\forall \theta}p(s|\theta)\cdot p(\theta|S)d\theta
    \label{equ:integral}
\end{equation}

The latter is no more than an expectation of a given parametric function, in this case Gaussian, with a parameter distribution, $p(\theta|S)$ that depends on our data set $S$. For this point we should focus all our attention on the second factor of the product, that is, obtaining a parameter distribution from $S$. We can do this by applying Bayes' theorem and defining a prior knowledge over the parameters. Thus:

\begin{equation}
    p(\theta|S)=\frac{p(S|\theta)\cdot p(\theta)}{p(S)}
\end{equation}

Where $p(S)$ is the area under the numerator function in the RHS of the equation. From this point we shall make some assumptions for having an analytic closed form for $p(s|S)$. First is that our model $p(s|\theta)$ is Gaussian. This means, taking in account that data from class $H$ only give information about that class and that samples are drawn iid, that we can factorize $p(S|\theta)$ as we have already done in Eq. \ref{equ:fact}. Our last assumption is that the prior over the parameters $p(\theta)$ is a Gaussian-gamma function and thus we can rewrite the expression as the conjugate prior $p(\theta|\phi)$ where hyperparameters $\phi$ govern the form of the prior knowledge about this function. The Gaussian-gamma function is a conjugate prior of the Gaussian function and therefore the posterior of the parameter is also Gaussian-gamma. The integral from Eq. \ref{equ:integral} has an analytic form which is a Student's t. We can find the details about the inference process in \cite{Minka} and \cite{Brumer}, where different approaches of the problem yield the same result. \cite{Minka} uses a non-informative parameter prior when the entropy of the parameter is maximum, and \cite{Brumer} uses a Gaussian gamma. By setting the parameters of the Gaussian-gamma to specific values we can have a Gaussian-gamma that tends to be non-informative. In our work we use the approach in \cite{Brumer}, assuming maximum entropy for the parameters of the model.

\begin{figure}[H]
    \centering
    \includegraphics[scale=0.30,trim={3cm 1cm 3cm 1cm}]{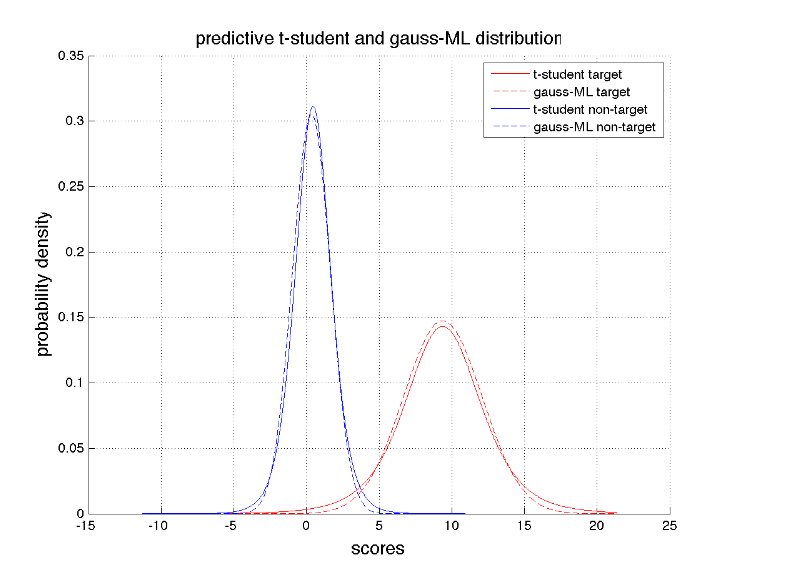}
    \caption{Likelihood Distributions. Dashed line represent ML densities and continuous line represent BI densities.}
    \label{fig:1}
\end{figure}

\begin{figure}[H]
    \centering
    \includegraphics[scale=0.30,trim={3cm 1cm 3cm 1cm}]{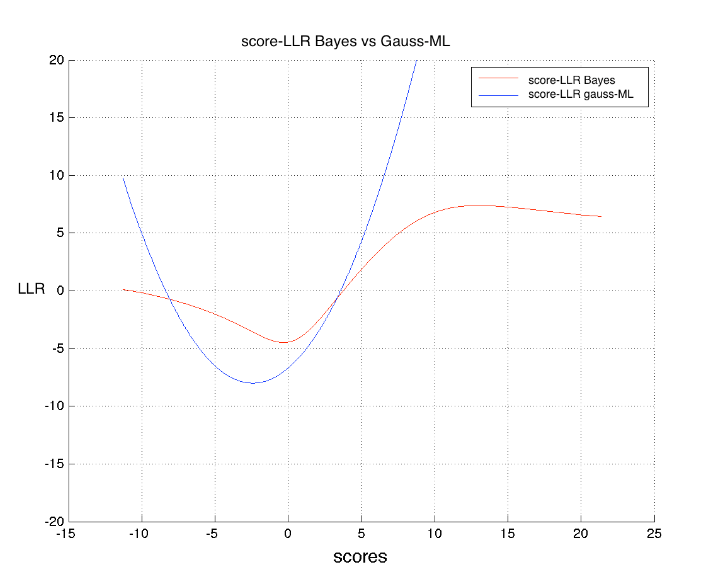}
    \caption{Function that transforms scores into log-likelihood-ratios (LLRs) for a given training dataset of scores.}
    \label{fig:my_label}
\end{figure}

When we have data sparsity, Bayesian inference (BI) (fully-Bayesian, as it is implemented here) is better than ML. When the number of data samples is big, the BI inference tends to the same result as ML. We see this effect reflected in the tails of the Student's t distribution. We observe that in the data space where we have enough samples we have a good likelihood fit, but with sparse data the tails of the Student's t are heavier. This is reflected in the likelihood ratio giving a preference for one class in the well represented data space and no preference in places without representation. Figures \ref{fig:1} and \ref{fig:my_label} show BI vs ML inference of the parameters. First we show the likelihood distributions and then the function that transforms scores into log-likelihood-ratios (LLRs) for a given dataset. It can be seen that the LLRs obtained by the fully-Bayesian model (BI) are much more limited when the scores become extreme. This is a desired effect in forensic science, because very big LR values make little sense for automatic speaker recognition systems. However, due to data sparsity, extremely big values of the LR are allowed with ML, which is a quite undesirable effect.

\section{\textsc{Experiments}}
\label{sec__experiments}

\subsection{Database and automatic speaker recognition system}
\label{ssec__databaseSystem}

For the experiments in this work, we used the data provided by NIST for the 2014 Speaker Recognition i-Vector challenge\footnote{https://www.nist.gov/sites/default/files/documents/itl/iad/mig/sre-ivectorchallenge\_2013-11-18\_r0.pdf}. For this challenge, 600-dimensional i-Vectors were provided from conversational telephone speech data available for previous NIST Speaker Recognition Evaluations (SRE's), from 2004 to 2012. Different amounts of speech were used to compute the i-Vectors, following a log normal distribution with mean of 39.58 seconds. From these i-Vectors, scores were generated using PLDA \cite{matejka11PLDA}. To develop the PLDA system, we used i-Vectors from utterances with more than 30 seconds of speech in the development set, and the ground truth labels provided by NIST after the evaluation. This subset consists of 17424 i-Vectors, from 3769 different speaker identities. The evaluation data provided for this challenge comprises five i-Vectors for each target speaker model, and single i-Vectors representing test segments. The number of target speaker models was 1306, and the number of test i-Vectors, 9634, resulting in over 12 million trials.

\subsection{Experimental Protocol}
For the experiments in this work, we have used the scores generated in the i-Vector challenge in different ways depending on the proposed anchoring scheme, as described below:

\begin{enumerate}
    \item Suspect-Anchored (SA): 
    \begin{itemize}
        \item $S_p$ is generated by drawing scores from a pool of scores including all possible combinations of two utterances from the suspect, without including the speech segment(s) of the suspect that might be present in the case. Different amounts of $N_p$ scores are drawn, in order to simulate data sparsity in $S_p$.
        \item $S_d$ is generated by comparing all utterances from the suspect with all utterances from other identities, excluding all utterances present in the case.
    \end{itemize}
    \item Reference-Anchored (RA): 
    \begin{itemize}
        \item $S_p$ is generated by drawing scores from a pool of scores $\Delta \left( q_p^{(i)},r\right)$, where $r$ is the reference suspect speech in the case, and $\left\{q_p^{(i)}\right\}$ are the remaining utterances from the suspect, excluding $q$ when it comes from the suspect. Different amounts of $N_p$ scores are drawn, in order to simulate data sparsity in $S_p$.
        \item $S_d$ is generated by comparing $r$, \emph{i.e.} the reference suspect speech in the case, with a number of speech segments from other identities, excluding all utterances present in the case.
    \end{itemize}
\end{enumerate}

Notice that, since score sparsity rarely affects $S_d$, $N_d$ is not constrained in order to focus on the sparsity effects in $S_p$. Finally, in order to use the same speech data for every value of $N_p$, only suspects with more than $10$ utterances have been selected. All in all, to measure performance, a number of $18192$ same-origin and $11560$ different-origin LR values have been computed as a minimum, with even more values depending on the value of $N_p$ and the anchoring scheme.

\begin{figure}[t!]
    \centering
    \includegraphics[scale=0.35,trim={3cm 1cm 3cm 1cm}]{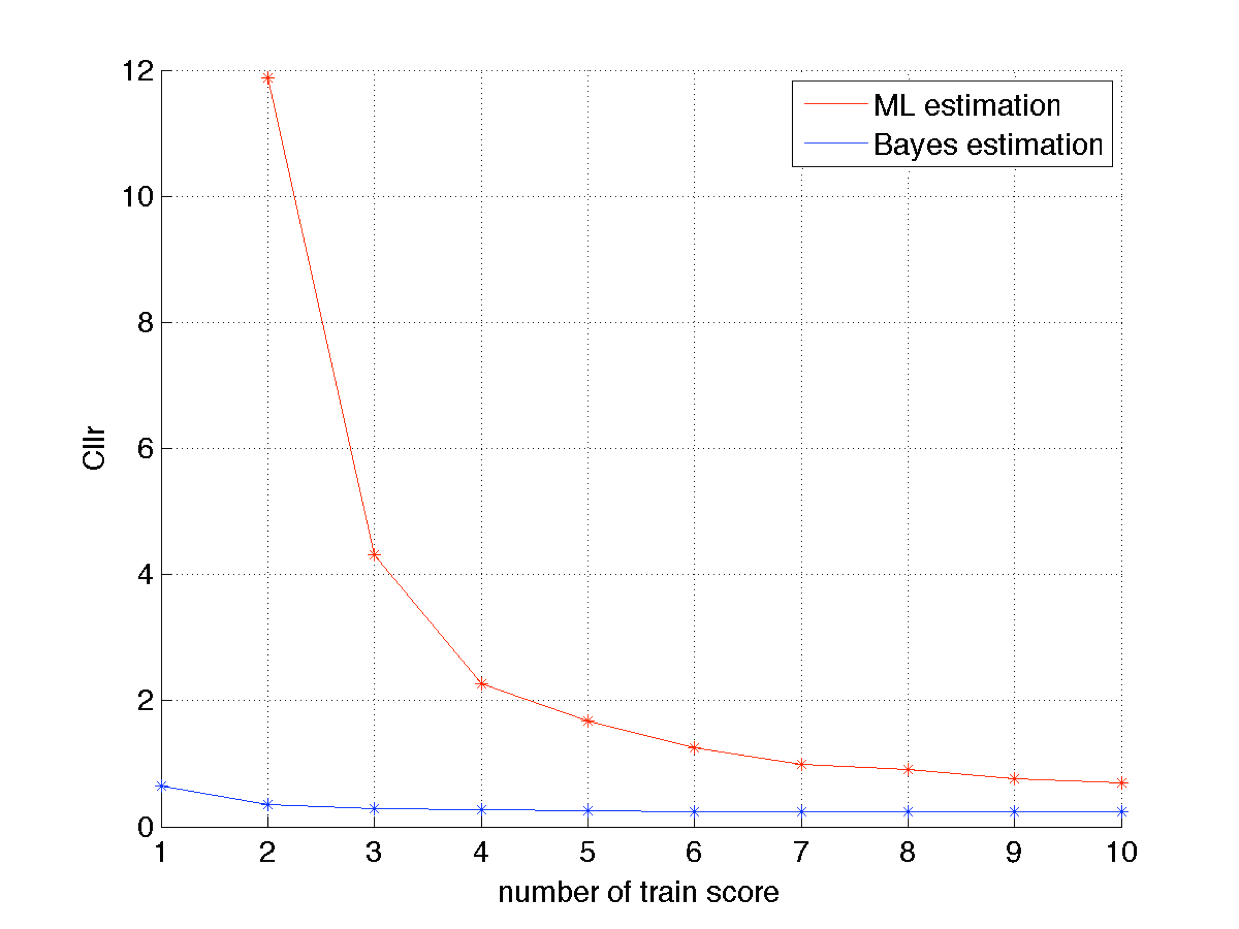}
    \caption{$C_{llr}$ values for the ML and fully-Bayesian models for the suspect-anchored (SA) scheme, as a function of $N_p$.}
    \label{fig__results_ML_vs_Bayes_SuspectAnchoring}
\end{figure}

\begin{figure}[t!]
    \centering
    \includegraphics[scale=0.35,trim={3cm 1cm 3cm 1cm}]{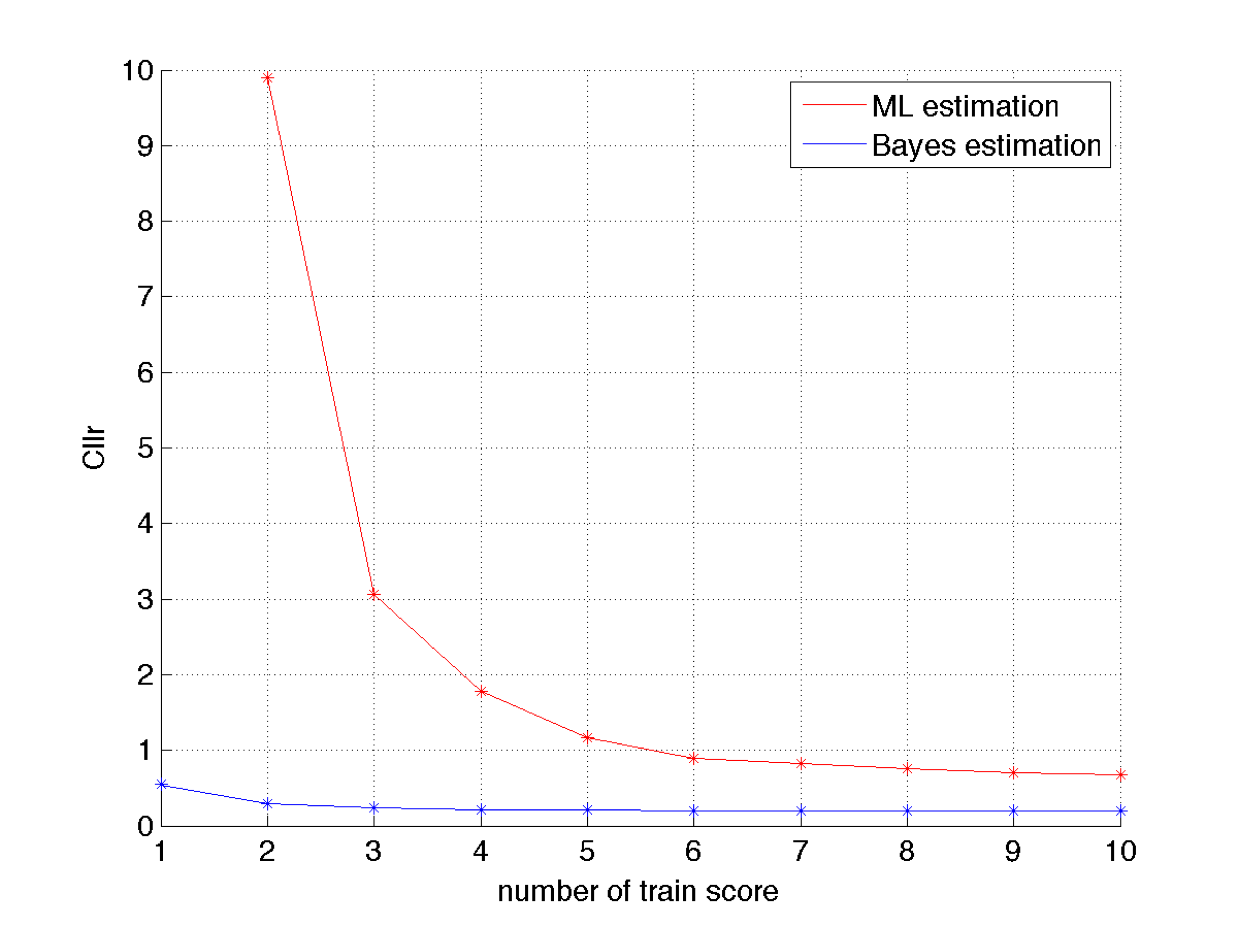}
    \caption{$C_{llr}$ values for the ML and fully-Bayesian models for the reference-anchored (SA) scheme, as a function of $N_p$}
    \label{fig__results_ML_vs_Bayes_ReferenceAnchoring}
\end{figure}

\subsection{Results}
Figure \ref{fig__results_ML_vs_Bayes_SuspectAnchoring} shows the results in terms of $Cllr$ of the ML and fully-Bayesian schemes in the Suspect-Anchored (SA) scheme; and Figure  \ref{fig__results_ML_vs_Bayes_ReferenceAnchoring} shows the corresponding results for the Reference-Anchoring (RA) scheme. Results show that the fully-Bayesian approach clearly outperforms ML for all sizes of the training score set $N_p$. Also, the value of $C_{llr}$ is much lower for this model, especially for lower values of $N_p$. This supports the hypothesis that the fully-Bayesian model allows the incorporation of uncertainty, due to score sparsity, to the LR in an effective way. This is especially relevant for the lowest values of $N_p$, because the ML approach yields much higher values of $C_{llr}$, which means much poorer performance. Moreover, $C_{llr}<1$ always obtains for the fully-Bayesian model, which means that the approach is always informative for evidence evaluation. This is not the case of ML, since $C_{llr}$ is much bigger than $1$ for many values of $N_p$.

Moreover, the comparison of both figures shows that the RA approach outperforms SA. This was expected, since RA allows the training scores to resemble the conditions of the score in each case in a much adequate way. 

\section{\textsc{Conclusions}}
\label{sec__conclusions}

In this work, the use of fully-Bayesian Gaussian models have proven to be adequate for forensic voice comparison using automatic systems. This is particularly true when compared to widely used ML models, that have shown to be very sensitive to sparsity in the training scores, a situation that often happens to training scores under $H_p$. On the other hand, fully-Bayesian methods effectively cope with the lack of data by incorporating the associated uncertainty, leading to much more moderate LR values, and drastically improving the value of $C_{llr}$. In fact, performance with fully-Bayesian models is always better than not reporting the $LR$ (meaning $C_{llr}<1$ always). Given the difficulty of the NIST i-Vector Challenge task, we can recommend fully-Bayesian methods to compute LRs in forensic practice, although more research is needed to compare the proposed approach with other models.

One interesting issue in this article is related to the anchoring schemes. We have reported experiments where a suspect-anchored and reference-anchored yields adequate (\emph{i.e.}, informative) and robust performance, the latter outperforming the former. Although this result confirms the research hypotheses, recent discussion on this topic motivates further research \cite{hepler11handwritingAnchoring,alberink14fingermarkConditioning}.

%\section{\textsc{Acknowledgements}}
%This work has been supported by proyect TEC2015-68172-C2-1-P from Spanish \emph{Ministerio de Economía y Competitividad}.

%\bibliography{references}
\bibliographystyle{elsarticle-num}

%\bibliography{references}
\end{document}